\documentclass[]{article}

\pdfoutput=1

\usepackage[authoryear]{natbib}
\usepackage{booktabs}  
\usepackage{amsfonts}
\usepackage{amsmath}
\usepackage{dcolumn}
\usepackage{bm}
\usepackage{cancel}
\usepackage{verbatim}
\usepackage{ifthen}
\usepackage{url}
\usepackage{sectsty}
\usepackage{balance}
\usepackage{graphicx} 
\usepackage{lastpage}
\usepackage[format=plain,justification=RaggedRight,singlelinecheck=false,font=small,labelfont=bf,labelsep=space]{caption}
\usepackage{fancyhdr}
\usepackage{subcaption}
\usepackage{makecell}
\usepackage{multirow}
\usepackage{fancyvrb}
\usepackage[ruled,vlined]{algorithm2e}
\usepackage{authblk}
\usepackage[letterpaper, margin=1in]{geometry} 

\usepackage{setspace}
\usepackage{cellspace}
\newcommand{\Tra}{^{\sf T}} 
\newcommand{\Inv}{^{-1}} 
 
\newcommand{\V}[1]{{\bm{\mathbf{\MakeLowercase{#1}}}}} 
\newcommand{\M}[1]{{\bm{\mathbf{\MakeUppercase{#1}}}}} 

\def\B{{\bf B}} 

\def\b{{\bf b}}

\def\D{{\bf D}}

\def\I{{\bf I}}

\def\R{{\bf R}}

\def\w{{\bf w}}

\def\X{{\bf X}} 
\def\x{{\bf x}}

\def\Y{{\bf Y}}
\def\y{{\bf y}} 

\def\Z{{\bf Z}} 
\def\z{{\bf z}}

\def \vmu{{\V{\mu}}}

\def \mpsi{{\M{\Psi}}} 
\def \vpsi{{\V{\psi}}}

\def \msigma{{\M{\Sigma}}}

\def \mlambda{{\M{\Lambda}}}

\def \mtheta{{\M{\Theta}}} 
\def \vtheta{{\V{\theta}}}

\def \vgamma{{\V{\gamma}}}

\newcommand{\normal}{\mathcal{N}}

\let\code=\texttt

\DefineVerbatimEnvironment{Sinput}{Verbatim}{fontshape=sl}
\DefineVerbatimEnvironment{Soutput}{Verbatim}{}
\DefineVerbatimEnvironment{Scode}{Verbatim}{fontshape=sl}

\DefineVerbatimEnvironment{Code}{Verbatim}{}
\DefineVerbatimEnvironment{CodeInput}{Verbatim}{fontshape=sl}
\DefineVerbatimEnvironment{CodeOutput}{Verbatim}{}

\let\proglang=\textsf
\newcommand{\pkg}[1]{{\fontseries{b}\selectfont #1}}

\providecommand{\keywords}[1]{\textbf{\textit{Keywords:}} #1}

\title{Divide-and-Conquer MCMC for Multivariate Binary Data}
\author[1]{Suchit Mehrotra}
\author[1]{Halley Brantley}
\author[1]{Peter Onglao}
\author[1]{Patricia Bata}
\author[1]{Roland Romero}
\author[1]{Jacob Westman}
\author[1]{Lauren Bangerter}
\author[2]{Arnab Maity}

\affil[1]{OptumLabs at United Health Group}
\affil[2]{Department of Statistics, North Carolina State University}

\date{}

\begin{document}

\maketitle

\doublespacing

\begin{abstract}
	
  The analysis of large scale medical claims data has the potential to improve
  quality of care by generating insights which can be used to create tailored
  medical programs.  In particular, the multivariate probit model can be used to
  investigate the correlation between multiple binary responses of interest in
  such data, e.g. the presence of multiple chronic conditions.  Bayesian
  modeling is well suited to such analyses because of the automatic uncertainty
  quantification provided by the posterior distribution. A complicating factor
  is that large medical claims datasets often do not fit in memory, which
  renders the estimation of the posterior using traditional Markov Chain Monte
  Carlo (MCMC) methods computationally infeasible.  To address this challenge,
  we extend existing divide-and-conquer MCMC algorithms to the multivariate
  probit model, demonstrating, via simulation, that they should be preferred
  over mean-field variational inference when the estimation of the latent
  correlation structure between binary responses is of primary interest.  We
  apply this algorithm to a large database of de-identified Medicare Advantage
  claims from a single large US health insurance provider, where we find
  medically meaningful groupings of common chronic conditions and asses the
  impact of the urban-rural health gap by identifying underutilized provider
  specialties in rural areas. 
	
\end{abstract}

\keywords{Big data, distributed computing, factor models, multivariate probit, 
parallel MCMC}

\section{Introduction}

Large scale medical data has become more accessible due to the advent of modern
database technology. Its analysis however, remains complicated by the fact that
these data sets do not fit in memory on a single machine, requiring most data
processing to be done on a massively parallel scale. This is not an issue when
calculating simple summary statistics, but is a substantial obstacle to fitting
many interesting statistical models. As interest in precision medicine grows,
the performance of big data algorithms for estimating parameters in statistical
models needs to be better understood, so that insights can be generated by
analyzing all the available data instead of subsets of the population.  

Developing algorithms for analyzing big datasets has been an active area of
research, with optimization based methods being the most common. In the
optimization literature, the most popular approach is to utilize stochastic
gradient descent (SGD) which, via software such as \pkg{TensorFlow}, can exploit
modern compute architecture to accelerate parameter estimation
\citep{tensorflow2015-whitepaper}. Alternatively, one can use the consensus ADMM
algorithm to estimate model parameters on multiple machines, but this approach
requires communication between each machine after each iteration of the
algorithm \citep{xu2017adaptive, brantley2020baseline}. In a frequentist
context, accurate uncertainty quantification via the bootstrap can be infeasible
for big data problems, since the model needs to be refit multiple times. 
\citet{kleiner2014scalable} propose the Bag of Little Bootstraps
to alleviate this issue, averaging the results of the bootstrap distribution
from a large number of subsamples from the original data set. For Bayesian
models, mean-field variational and stochastic variational inference algorithms
can be used to approximate the posterior distribution for large data sets, but
their performance, especially with respect to uncertainty quantification, can
deteriorate if parameters in the model are highly correlated
\citep{hoffman2013svi, vbreview}. Consequently, methods to scale Markov Chain
Monte Carlo (MCMC) algorithms, the gold standard for posterior approximation, to
large data sets is of paramount importance. 

For the purposes of this paper we restrict our attention to the situation where
we have data on a large number of individuals, but the parameter space of our
model is substantially smaller than the number of observations. This type of
data is common in large scale medical databases, where the number of individuals
available for analysis are an order of magnitude larger than potential
covariates, such as, lab results, diagnosis and procedure codes, and medication
information. Analyzing such data, dubbed `tall data', in a general class of
models has been an active area of research in the MCMC literature and we refer
the reader to \citet{bardenet2017markov} for a thorough review.
Like stochastic gradient descent optimization, gradient based samplers such as
Hamiltonian Monte Carlo \citep{betancourt2017conceptual} have been modified to
allow large datasets to be processed on a single machine, utilizing subsamples
of the data to transition between states \citep{welling2011bayesian,
ma2015complete}. This approach suffers from two major issues. First, the
requirement that a gradient be evaluated at each iteration of the algorithm
requires that discrete parameters be integrated out of the model. Second, like
SGD algorithms, stochastic approaches to gradient based MCMC require the data
analyst to specify step sizes for the algorithm; these function like tuning
parameters, and finding optimal values for them can increase the total
computational cost of fitting the model. \citet{baker2019sgmcmc}
aim to make stochastic gradient MCMC accessible to applied researchers via the
\proglang{R} package \pkg{sgmcmc}, which utilizes \pkg{TensorFlow} to
automatically calculate gradients for subsamples of the data. 

While stochastic gradient MCMC can be difficult to implement, divide-and-conquer
MCMC provides an embarrassingly parallel alternative to scale any MCMC algorithm
to a tall data setting. This approach divides the dataset into disjoint subsets
(shards), runs a MCMC algorithm on each shard, and combines these subset
posterior samples to arrive at a final posterior distribution. The main question
for this line of research has been finding optimal combination strategies for
the subset posteriors. \citet{huang2005sampling} utilize
Gaussian approximations for each subset posterior, which become more accurate as
the data size in each shard increases. \citet{consensusmc} propose
a similar approach, but instead of multiplying resulting posteriors, they take a
weighted average of each sample, which is exact if each subset posterior is
Gaussian. \citet{neiswanger2013asymptotically} propose a
non-parametric and semi-parametric approach to combining the posteriors, and
show that their method is asymptotically correct as the number of samples from
each subset posterior goes to infinity. \citet{minsker2014scalable} calculate the 
geometric median of subset posteriors,
while \citet{srivastava2015wasp} combine the posterior by
calculating their Wasserstein barycenter (WASP). Finally, \citet{li2017simple} 
approximate the posterior by averaging the estimated
quantiles from each subset posterior.  

A major advantage of the divide-and-conquer approach is that it can be combined
with other methods for dealing with difficult Bayesian problems. For example, in
a regression model with both a large number of subjects and predictors, one
could use GPU accelerated Gibbs sampling for each subset, thereby making full
use of a GPU cluster \citep{terenin2019gpu}. Therefore, we think that the
divide-and-conquer approach to MCMC will be an important tool for applying
complex Bayesian models to large datasets, even as improvements continue to be
made in algorithms that only need a single machine to perform MCMC for big data. 

In this paper, we use divide-and-conquer MCMC to fit multivariate probit models
on a de-identified data set of over three million Medicare Advantage members in
a research database from a single large US health insurance provider (the
UnitedHealth Clinical Discovery Database). Using this approach, we are able to
address two important questions in the medical literature. 

First, we investigate the relationships between chronic conditions in the
Medicare Advantage population to assist in the creation of therapies and
programs designed to address co-occurring diseases. Finding patterns among
individuals with multiple comorbidities (multimorbidity) can help providers and
insurers improve quality of care by tailoring programs to the needs of specific
groups. Much work exists in the current literature which aims to find subgroups
within patient populations via clustering. For example, 
\citet{newcomer2011identifying} cluster individuals who had retrospectively been
in the top 20\% in cost during a particular year. To find distinctive groups
they utilized agglomerative hierarchical clustering on approximately 20 chronic
conditions. For a thorough review of the literature, we refer the reader to
\citet{prados2014multimorbidity}, who conducted a
literature review to identify patterns in multimorbidity, and
\citet{violan2014prevalence}, who conducted another literature review, but
focused on predicting multimorbidity using demographic characteristics. 

Second, we explore the disparities in healthcare utilization between rural and
urban individuals. Rural Americans have a mortality disadvantage relative to
their urban counterparts which is driven in part by having poorer access to
medical care \citep{weisgrau1995issues, james2017long}. Rural individuals have
lower access to doctors; for example, \citet{hing2014state} show
that rural areas have approximately 40 physicians per 100,000 people, compared
to approximately 55 for urban and suburban areas. Rural residents often have
long wait times to access specialty care and have to travel long distances to
get to their appointments \citep{cyr2019access}. The wide adoption of
telemedicine during the COVID-19 pandemic provides the US health system with an
opportunity to address some of these care discrepancies, and insight is needed
to identify medical specialties that would provide the greatest benefits to
rural individuals.

To the best of our knowledge, we have not found any analyses in the statistics
or medical literature applying the multivariate probit model to a data set of
this size. It also seems that the medical literature primarily utilizes multiple
univariate linear models to individually analyze correlated endpoints of
interest. Consequently, this paper adds to the existing literature by exploring
the behavior of divide-and-conquer MCMC for the multivariate probit model and by
providing use cases of how it can be applied to answer important medical
questions. The rest of this paper proceeds as follows: we review the
multivariate probit model in Section \ref{sec:mvp:mvp}, discuss approaches for
combining subset posteriors in Section \ref{sec:mvp:consensus}, conduct a
simulation study comparing stochastic variational inference with
divide-and-conquer MCMC in Section \ref{sec:mvp:sims}, discuss applications in
Section \ref{sec:mvp:applications}, and conclude the paper with Section
\ref{sec:mvp:discuss}.

\section{Multivariate Probit Model} \label{sec:mvp:mvp}

Let $\y_n$ be a $M \times 1$ dimensional vector of binary responses for
individual $n \in \{1, \dots, N\}$. The multivariate probit model utilizes a
multivariate Gaussian latent variable, $\z_n$, to connect the binary data to a
continuous latent space where the underlying covariance structure between the
responses can be modeled \citep{ashford1970multi, chibmvp}. Letting $\vmu_n$ be
a mean vector that can be dependent on covariates, we can write the relationship
of the observed response to the underlying latent variables as: 
\begin{align}
\begin{split} \label{eq:mvp:full_cov}
	y_{nm} & = \mathbb{I}(z_{nm} > 0), \\
	\z_n & \sim \normal \left( \vmu_n, \msigma \right). \\
\end{split} 
\end{align}
Integrating over the latent variables gives us the likelihood of the observed
data: 
\begin{align} \label{eq:mvp:cov_prob}
	P(\Y_n = \y_n) = \int_{R(y_{n1})} \cdots \int_{R(y_{nM})} (2 \pi)^{-\frac{M}{2}} 
	|\msigma|^{-\frac{1}{2} } \exp \left\{ 
		-\frac{1}{2} (\z_n - \vmu_n)\Tra \msigma\Inv (\z_n - \vmu_n)
	\right\} d \z_n,
\end{align}
where $R(y_{nm}) = (-\infty, 0]$ if $y_{nm} = 0$ or $(0, \infty)$ if $y_{nm} =
1$. The probability in \eqref{eq:mvp:cov_prob} remains the same even if we
rescale the parameters to a correlation scale \citep{tabet2007bayesian,
lawrence2008bayesian, taylor2017}. That is, letting $\D = diag(\msigma)$ and
setting $\R = \D^{-\frac{1}{2}} \msigma \D^{-\frac{1}{2}}$, $\tilde{\vmu}_n =
\D^{-\frac{1}{2}} \vmu_n$, and $\tilde{\z}_n = \D^{-\frac{1}{2}} \z_n$, it can
be shown that \eqref{eq:mvp:cov_prob} is equivalent to: 
\begin{align*}
	P(\Y_n = \y_n) = \int_{R(y_{n1})} \cdots \int_{R(y_{nM})} (2 \pi)^{-\frac{M}{2}} 
		|\R|^{-\frac{1}{2} } \exp \left\{ 
	-\frac{1}{2} (\tilde{\z}_n - \tilde{\vmu}_n)\Tra \R\Inv (\tilde{\z}_n - \tilde{\vmu}_n)
	\right\} d \tilde{\z}_n.
\end{align*}
This implies that the model is only identifiable if we fix the diagonals of the
covariance matrix, which we constrain to one to estimate the model on a
correlation scale. Fortunately, the lack of identifiability of the parameters is
not an issue during sampling. To generalize, if we let $\vmu_n = \B \x_n$, where
$\B$ is a $M \times P$ matrix of regression coefficients and $\x_n$ is a $P
\times 1$ vector of predictors, and utilize conjugate priors for $\B$ and
$\msigma$, we can estimate the model parameters with a Gibbs sampler. After each
iteration of the algorithm, we rescale the parameters to the correlation scale
by implementing the transformations above, and store these transformed values as
samples from the posterior distribution \citep{mcculloch1994exact,
lawrence2008bayesian, taylor2017}. 

The formulation in \eqref{eq:mvp:full_cov} has a few computational bottlenecks.
First, sampling a covariance matrix is a computationally intensive task and is
unstable in high dimensions. Additionally, the full conditional distribution for
$\z_n$ is a truncated multivariate normal distribution, which is also
computationally intensive to sample from in high dimensions. Consequently, for
the purposes of this chapter, we utilize a factor model for the covariance
structure, a popular approach to bypass the bottlenecks discussed above
\citep{hahn2012sparse, taylor2017}. We also utilize non-informative priors for
the regression coefficients and the factor matrix, which leads to the hierarchy: 
\begin{align}
\begin{split} \label{eq:mvp:factor_cov}
	y_{nm} & = \mathbb{I}(z_{nm} > 0), \\
	\z_n & \sim \normal \left(
	 	\B \x_n\Tra + \mtheta \vpsi_n, \mlambda 
	\right), \\
	\b_{m} & \sim \normal \left( \V{0}, 10^6 \I \right), \\
	\vtheta_{m} & \sim \normal \left( \V{0}, 10^6 \I \right), \\
	\vpsi_n & \sim \normal \left( \V{0}, \I \right),
\end{split} 
\end{align}
where $\b_m$ is the $m^{th}$ row of $\B$, $\mtheta$ is a $M \times K$ matrix,
with $\mtheta = (\vtheta_1, \dots, \vtheta_M)\Tra$, and $\vpsi_n$ is a $K \times
1$ vector.

Integrating out $\vpsi_n$ from \eqref{eq:mvp:factor_cov} implies that $cov(\z_n)
= \mtheta \mtheta\Tra + \mlambda$. It is well known that this representation for
the covariance matrix does not have a unique solution in $\mtheta$ unless
constraints on its form are implemented. Fortunately, since our primary interest
is estimation of the correlation between the responses, we can, like with the
covariance matrix, sample $\mtheta$ without restrictions \citep{dunsonmgp}.
Additionally, following \citet{hahn2012sparse}, we set $\mlambda =
\I$ which implies that the probability of a particular response equaling one is:
\begin{align*}
	P(y_{nm} = 1) = \Phi(\x_n\Tra \b_m + \vpsi_n\Tra \vtheta_m), 
\end{align*}
where $\Phi$ is the CDF of a univariate standard normal distribution. The
parameters in \eqref{eq:mvp:factor_cov} can be estimated via a Gibbs sampler and
at each iteration of the algorithm we calculate $\msigma = \mtheta \mtheta\Tra +
\I$, $\D = diag(\msigma)$, and store $\R = \D^{-\frac{1}{2}} \msigma
\D^{-\frac{1}{2}}$ and $\tilde{\B} = \D^{-\frac{1}{2}} \B$, as samples from the
posterior distribution \citep{taylor2017}. Letting $\mpsi$ be an $N \times K$
matrix with $\mpsi = (\vpsi_1, \dots, \vpsi_N)\Tra$, the Gibbs sampler for
estimating the parameters in \eqref{eq:mvp:factor_cov} is given in Algorithm
\ref{algo:mvp:gibbs-sampler}. 

\begin{algorithm}[t]
	\SetAlgoLined
	\KwResult{Samples for regression coefficients, $\tilde{\B}$, and correlation matrix, $\R$.}
	\KwIn{Number of factors ($K$), \code{n\_iter}, \code{burn\_in}, $\X_{N \times P}$, and $\Y_{N \times M}$}
	Initialize $\mpsi_{N \times K}$, $\mtheta_{M \times K}$, and $\B_{M \times P}$ \;
	\For{$i\gets1$ \KwTo \code{n\_iter}}{
		\For{$n\gets1$ \KwTo \code{N}}{
			\For{$m\gets1$ \KwTo \code{M}}{
				\eIf{$y_{nm} = 1$}{
					$z_{nm} \sim 
					\mathcal{TN}_{+}(\x_n\Tra \b_m + \vpsi_n\Tra \vtheta_m, \ 1)$
				}{
					$z_{nm} \sim 
					\mathcal{TN}_{-}(\x_n\Tra \b_m + \vpsi_n\Tra \vtheta_m, \ 1)$
				}
				$
				\vpsi_n \sim \normal \left\{
				(\mtheta\Tra \mtheta + \I)\Inv \mtheta\Tra (\z_n - \B \x_n), \
				(\mtheta\Tra \mtheta + \I)\Inv 
				\right\} 
				$ \;
		}}
		\For{$m\gets1$ \KwTo \code{M}}{
			$
			\b_m \sim \normal \left\{
			\left( \X\Tra \X + 10^{-6} \I \right)\Inv
			\X\Tra \left(\z_m - \mpsi \vtheta_m \right), \ 
			\left( \X\Tra \X + 10^{-6} \I \right)\Inv
			\right\} $ \;
			$
			\vtheta_m \sim \normal \left\{ 
			\left( \mpsi\Tra \mpsi + 10^{-6} \I \right)\Inv 
			\mpsi\Tra  \left(\z_m - \X \b_m \right), \
			\left( \mpsi\Tra \mpsi + 10^{-6} \I \right)\Inv 
			\right\} 
			$ \;
		}
		$\msigma = \mtheta \mtheta\Tra + \I$ \; 
		$\D = diag(\msigma)$ \;
		$\R^{(t)} = \D^{-\frac{1}{2}} \msigma \D^{-\frac{1}{2}}$ \;
		$\tilde{\B}^{(t)} = \D^{-\frac{1}{2}} \B$ \;
	}
	\KwOut{$\left[
		\left\{
		\tilde{\B}^{(\code{burn\_in} + 1)}, \cdots,
		\tilde{\B}^{(\code{n\_iter})}
		\right\}, 
		\left\{
		\R^{(\code{burn\_in} + 1)}, \cdots,
		\R^{(\code{n\_iter})}
		\right\}
		\right]$}
  \caption{Gibbs Sampler for the Multivariate Probit Model
  \eqref{eq:mvp:factor_cov}} \label{algo:mvp:gibbs-sampler}
\end{algorithm}

\section{Divide and Conquer MCMC} \label{sec:mvp:consensus}

Each iteration of the Gibbs sampler in Algorithm \ref{algo:mvp:gibbs-sampler}
needs to cycle through the whole data set and sample a latent variable, $\z_n$,
for each data point, $\y_n$, which becomes a computational bottleneck when $N$
is large. Additionally, as the number of data points increases, loading the full
dataset into memory can be problematic. Bypassing these issues is possible in an
optimization framework via the utilization of stochastic gradient descent (SGD)
algorithms. Stochastic variational inference (SVI) uses stochastic gradient
descent combined with mean-field variational inference to the approximate
posterior distribution, but these methods are known to be inadequate for
uncertainty quantification \citep{hoffman2013svi, vbreview}. However, when it
comes to MCMC methods, no clear guidance exists on the appropriate methodology
to use for datasets that do not fit in memory \citep{bardenet2017markov}. In
this section, we review one of the simplest approaches for scaling any MCMC
algorithm to data sets where each data point has a corresponding latent
variable. This approach splits the data into disjoint subsets (shards), runs
a MCMC algorithm on each shard, and uses a combination strategy to arrive at a
final posterior distribution. This is a principled approach to MCMC since the
posterior distribution of a general parameter, $\vgamma$, can be written as a
product of the posterior distributions of disjoint subsets:
\begin{align*}
p(\vgamma | \w) \propto p(\w | \vgamma) p(\vgamma) = 
    \prod_{s = 1}^{S} p(\w_s | \vgamma) p(\vgamma)^{\epsilon_s},
\end{align*}
where $\w$ is a vector containing all observations and $\epsilon_s$ is the
proportion of the data in shard $s$, with $\w_s$ being the observations in the
shard. 

This divide-and-conquer approach can easily be applied to the Gibbs sampler in
Algorithm \ref{algo:mvp:gibbs-sampler}. In this situation, the full posterior
distribution of the parameters can be written as: 
\begin{align}
	p(\Z, \mpsi, \mtheta, \B | \Y, \X) & \propto \prod_{s = 1}^S 
		p(\Y_s | \Z_s) 
		p(\Z_s | \mpsi_s, \X_s, \mtheta, \B) 
		p(\mpsi_s) 
		p(\mtheta, \B)^{\epsilon_s}
\end{align}
where $\Y_s$, $\X_s$, $\Z_s$, and $\mpsi_s$ are the parameters restricted to
those in $s^{th}$ shard. As can be seen, the priors for $\mtheta$ and $\B$ are
raised to the $\epsilon_s$ power, which only necessitates that the Gibbs sampler
be modified by adding an $\epsilon_s$ to the prior component of the update in
Algorithm \ref{algo:mvp:gibbs-sampler}. That is, the update for each $\b_m$ and
$\vtheta_m$ in shard $s$ becomes:
\begin{align*}
\b_m & \sim \normal \left\{
	\left( \X_s\Tra \X_s + \epsilon_s 10^{-6} \I \right)\Inv
		\X_s\Tra \left(\z_m - \mpsi_s \vtheta_m \right), \ 
	\left( \X_s\Tra \X_s + \epsilon_s 10^{-6} \I \right)\Inv
\right\}, \\
\vtheta_m & \sim \normal \left\{ 
	\left( \mpsi_s\Tra \mpsi_s + \epsilon_s 10^{-6} \I \right)\Inv 
		\mpsi_s\Tra  \left(\z_m - \X_s \b_m \right), \
	\left( \mpsi_s\Tra \mpsi_s + 10^{-6} \I \right)\Inv 
\right\}.
\end{align*}
We would continue to store the identified parameters for each shard, letting
them be $\R^{(t)}_s$ and $\tilde{\B}^{(t)}_s$, and use a subset posterior
combination strategy on only the samples of the identifiable parameters. It
should also be noted that while we only use non-informative priors for $\b_m$
and $\vtheta_m$, the above discussion can be easily modified to account for
priors which promote a particular structure in either $\B$ or $\mtheta$. 

The primary computational roadblock to efficiently implementing
divide-and-conquer approaches is the computational complexity of the posterior
combination algorithm. It is especially important to be mindful of this in a
multivariate probit model since the number of parameters increase with both the
number of responses and predictors. Consequently, for the purposes of this
paper, we restrict our attention to the independent Consensus Monte Carlo (CMC)
\citep{consensusmc} and the posterior interval estimation (PIE)
\citep{li2017simple} algorithms due to their computational efficiency; they only
require taking averages. Alternatively, WASP  \citep{srivastava2015wasp}
requires solving a sparse linear program, which can be a computationally
intensive if proprietary optimization packages, such as Gurobi, are not readily
available, while the approach of
\citet{neiswanger2013asymptotically} requires an additional sampling step, which
is computationally intensive as the number of parameters increases.

\section{Simulations} \label{sec:mvp:sims}

In this section, we conduct a simulation study to compare the PIE and CMC
\citep{consensusmc, li2017simple} posterior combination algorithms with the
stochastic variational inference (SVI) algorithm implemented in the \proglang{R} 
package \pkg{vir} \citep{mehrotra2021variational}. We explore the performance 
of the algorithms as the number of
observations increases, $N \in \left\{10K, 20K, 40K, 60K, 80K, 100K \right\}$,
while fixing the number of responses, $M$, at 60, and the number of predictors,
$P$, at 50. We simulate $D = 15$ datasets from the hierarchy in
\eqref{eq:mvp:factor_cov}, using $L = 40$ factors for the covariance matrix, and
draw each element of $\b_0$, $\B$, $\X$, and $\mtheta$ from a $\normal(0, 1)$
distribution. We use the MSE and coverage of the regression coefficients, $\B$,
and the mean absolute error of the upper-triangular portion of the correlation
matrix to evaluate the algorithms. The formulas to calculate the metrics are
given below: 
\begin{align*} 
\begin{split}
	MSE & = \frac{1}{D M P} \sum_{d = 1}^D \sum_{m = 1}^M \sum_{p = 1}^P (b_{dmp} - \hat{b}_{dmp})^2 \\
	COV & = \frac{1}{D M P} \sum_{d = 1}^D \sum_{m = 1}^M \sum_{p = 1}^P
		\mathbb{I}(\hat{l}_{dmp} \leq b_{dmp} \leq \hat{u}_{dmp}) \\
	MAE & = \frac{1}{D [(M^2 + M) / 2 - M]} \sum_{d = 1}^D \sum_{i = 1}^M \sum_{j > i}^M |r_{dij} - \hat{r}_{dij}|
\end{split}
\end{align*}
where $\hat{b}$ and $b$ are the estimated and the true regression coefficients,
$\hat{l}$ and $\hat{u}$ are the lower and upper bounds of a 95\% credible
interval calculated using the quantiles of the posterior distribution, and $r$
and $\hat{r}$ are the true and estimated elements of the correlation matrix. 

\begin{figure}
	\centering
	\begin{subfigure}{\textwidth}
		\includegraphics[scale=0.575]{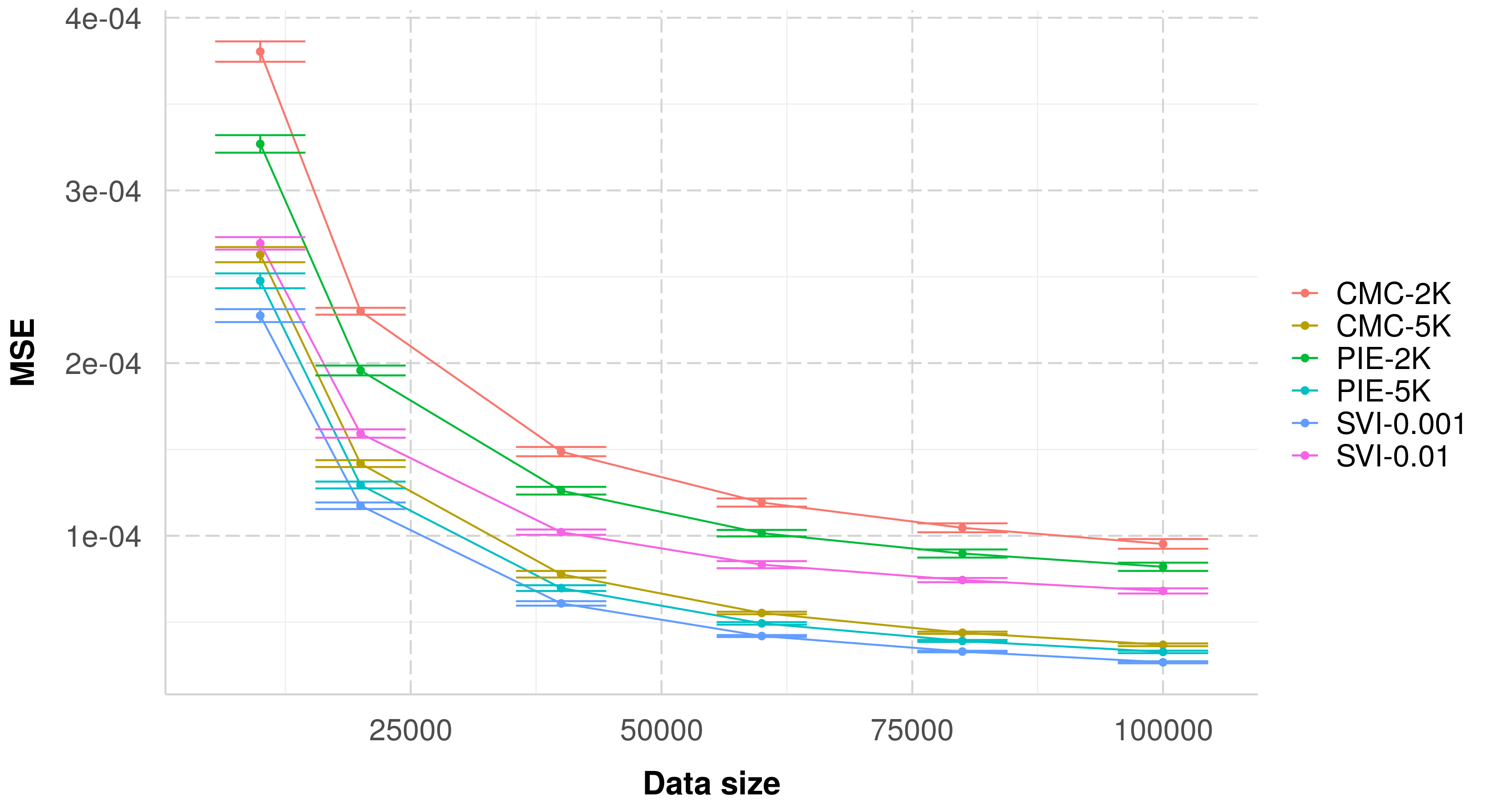}
	\end{subfigure}
	\begin{subfigure}{\textwidth}
		\includegraphics[scale=0.575]{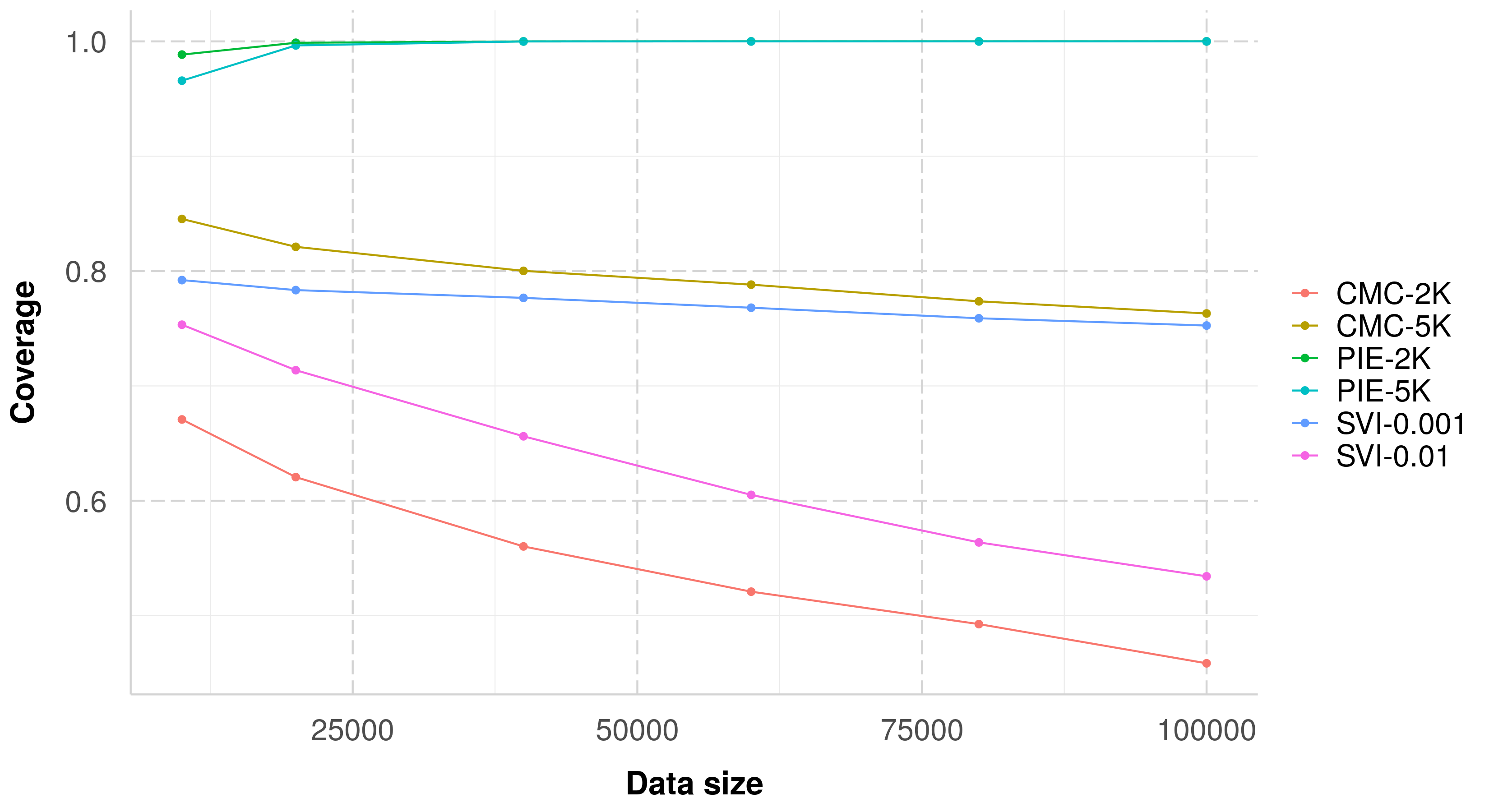}
	\end{subfigure}
  \caption{Simulation results for mean squared error (MSE) and coverage of the
  regression coefficients.}	\label{fig:cmc:b_mse}
\end{figure}

We hold the shard size constant for the posterior combination algorithms and
instead increase the number of shards as $N$ increases. For example, the CMC-2K
algorithms fix shard size at 2,000 data points and use 20 shards when the data
size is 40,000 and 50 shards when the data size is 100,000, whereas CMC-5K uses
8 and 20 shards, respectively. This setup mimics the real-world situation where
the limiting computational consideration is total run time and not the number of
available machines for running the MCMC algorithms. For the SVI algorithms, we
compare two constant step sizes of $0.001$ and $0.01$ and utilize a batch size
of 100 for each iteration. We run the MCMC algorithms for 15,000 iterations per
shard, using 10,000 of those iterations for burn in, and the SVI algorithms for
50,000 iterations. 

The results for MSE and coverage for the regression coefficients are given in
Figure \ref{fig:cmc:b_mse}. It can be seen that among the divide-and-conquer
approaches, the PIE algorithm outperforms the CMC algorithm. It should be noted
that, for both algorithms, better parameter estimation was achieved when shard
size was fixed at 5K instead of 2K; this is not surprising since the convergence
theory for both approaches requires that the size of each shard go to infinity.
In terms of the average MSE of the regression coefficients, the SVI algorithm
with step size $0.001$ outperforms all other algorithms; this provides support
for the fact that SVI algorithms are appropriate to use when point estimates for
the mean structure of the model are needed. In terms of coverage however, all
algorithms do not achieve 95\% coverage from 95\% credible intervals. The CMC
and SVI algorithms under-cover, while the PIE algorithm over-covers. For both
the CMC and SVI algorithms, we see decreasing coverage as the data size
increases, while for the PIE algorithms we see coverage go to one. We also
investigated the length of the credible sets and saw that the CMC and SVI
algorithms had intervals that narrowed as the sample size increased, whereas the
interval length for the PIE algorithm remained constant.

\begin{table}
  \caption{Mean absolute error (MAE) of the upper-triangular portion of the
  correlation matrix as data size increases. Average standard errors for the SVI
  algorithms are 0.0005, while the average standard errors for the
  divide-and-conquer approaches are 0.0002.} \label{tab:cmc:cor_mae} 
\begin{center}
\begin{tabular}{lrrrrrr}
\hline\hline
\multicolumn{1}{l}{}&\multicolumn{1}{c}{N = 10K}&\multicolumn{1}{c}{N = 20K}&\multicolumn{1}{c}{N = 40K}&\multicolumn{1}{c}{N = 60K}&\multicolumn{1}{c}{N = 80K}&\multicolumn{1}{c}{N = 100K}\tabularnewline
\hline
CMC-2K&$0.0215$&$0.0165$&$0.0132$&$0.0118$&$0.0111$&$0.0106$\tabularnewline
CMC-5K&$0.0181$&$0.0132$&$0.0096$&$0.0081$&$0.0072$&$0.0065$\tabularnewline
PIE-2K&$0.0203$&$0.0157$&$0.0128$&$0.0118$&$0.0112$&$0.0107$\tabularnewline
PIE-5K&$0.0175$&$0.0126$&$0.0091$&$0.0077$&$0.0069$&$0.0063$\tabularnewline
SVI-0.001&$0.1107$&$0.1108$&$0.1097$&$0.1109$&$0.1117$&$0.1105$\tabularnewline
SVI-0.01&$0.1109$&$0.1111$&$0.1097$&$0.1111$&$0.1120$&$0.1109$\tabularnewline
\hline
\end{tabular}\end{center}

\end{table}

Table \ref{tab:cmc:cor_mae} shows the mean absolute error of the upper
triangular elements of the correlation matrix for the respective algorithms. Of
note here is that the SVI algorithms perform poorly; the MAE does not improve as
sample size increases. This may be due to the fact that the large correlation
between $\vpsi_n$ and $\z_n$ is ignored in the mean-field assumption of the
algorithm. On the other hand, the CMC and PIE algorithms improve in performance
as the data size increases. As with the results for the regression coefficients,
we see that increasing the size of a shard leads to improved estimation of the
correlation matrix. Additionally, we also see that the PIE combination approach
performs slightly better than the CMC approach. 

In addition to the improved performance relative to CMC, PIE has an advantage
with regards to the memory needed to combine posteriors. Since PIE estimates the
posterior distribution by averaging quantiles, one can draw a large number of
samples from each shard yet only store a small number of quantiles at which
uncertainty quantification is of interest. Therefore, we use the PIE algorithm
in our applied analyses in Section \ref{sec:mvp:applications}.

\section{Applications} \label{sec:mvp:applications}

In this section we analyze a dataset of de-identified administrative claims for
Medicare Advantage members in a research database from a single large US health
insurance provider. The database contains medical claims for services submitted
for third party reimbursement, available as International Classification of
Diseases, Tenth Revision, Clinical Modification (ICD-10-CM). We use the ICD
codes to identify chronic conditions via version twenty two of the Centers for
Medicare \& Medicaid Services' (CMS) hierarchical condition categories (HCC), a
provider's specialty classification, and an individual's demographic information
in our analyses below. We show how the multivariate probit model
\eqref{eq:mvp:factor_cov}, combined with divide-and-conquer MCMC, can be used to
answer important questions using large medical claims data sets. First, we
investigate patterns of chronic conditions in individuals with multiple HCCs to
better identify areas of opportunity for targeted medical programs. Second, we
investigate the urban-rural divide in healthcare, by analyzing provider
utilization patterns of urban and rural individuals, relative to their suburban
counterparts.

\subsection{Co-morbid Conditions in the Medicare Advantage Population} \label{sec:mvp:medicare}

\begin{figure}
	\centering
	\includegraphics[scale=0.7]{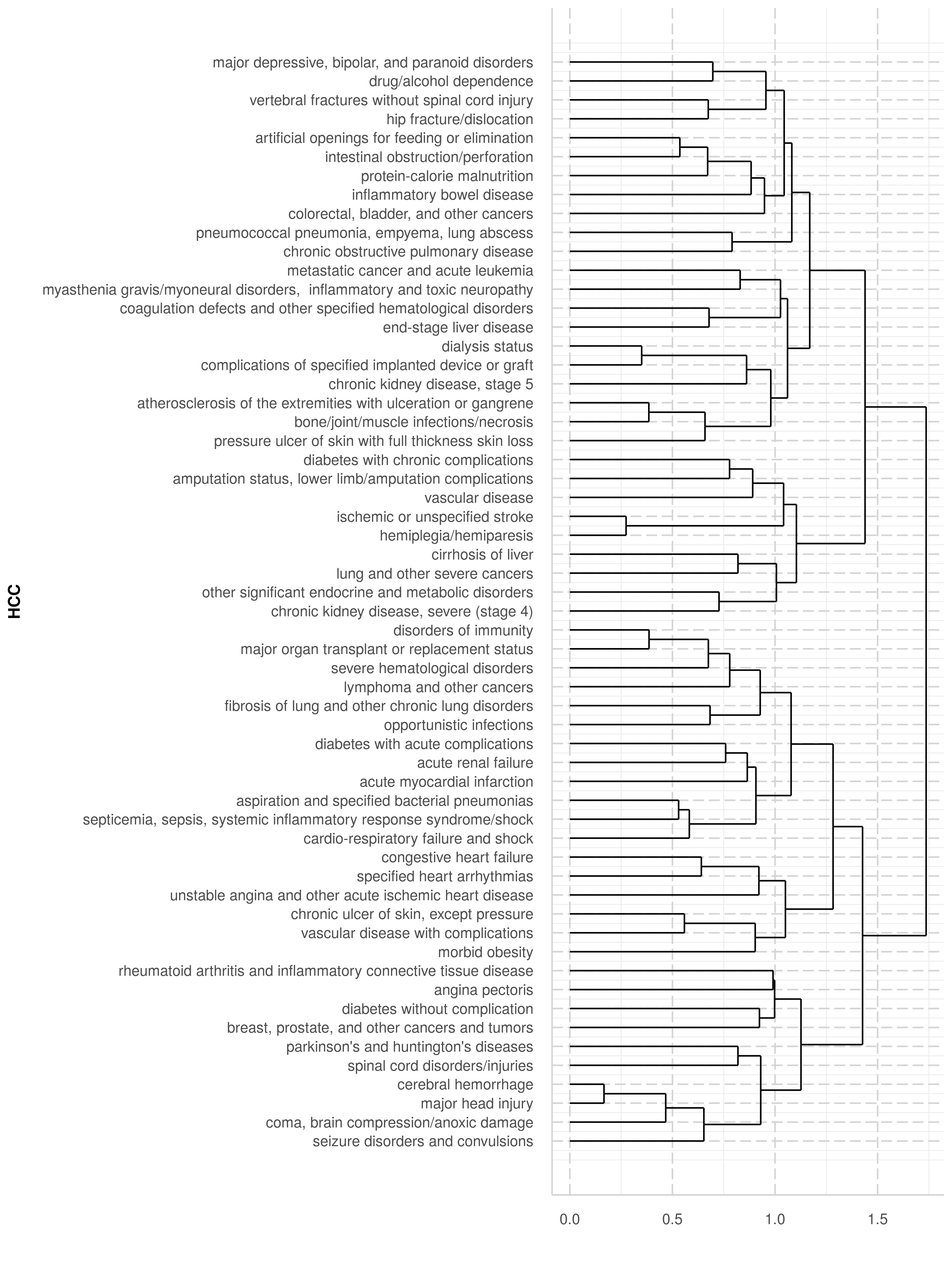}
  \caption{Dendrogram of the correlation matrix using one minus the correlation
  as a distance measure.}	\label{fig:mvp:dendro}
\end{figure}

We fit the multivariate probit model using an individual's HCCs in 2018 as the
response variable. As predictors, we use a member's age, gender, and a flag for
whether they were in a dual-eligible special needs plan (DSNP) during the year
as predictors; DSNP enrollees are those enrolled in both Medicare Advantage and
Medicaid. To find patterns in individuals who have multiple comorbidities, we
restrict our attention to individuals who were diagnosed with at least four HCCs
during the year and to the HCCs which were prevalent in at least one percent of
the restricted population; the list of HCCs considered is shown in Figure
\ref{fig:mvp:dendro}. The final data set had approximately 500,000 individuals,
58 HCCs, and three predictors. We fit the model using 10 shards, 40 factors, and
ran each sampler for 25,000 iterations, with 15,000 used for burn in. We
combined individual shards using the PIE algorithm, and used the median of the
correlation parameters as point estimates to visualize the correlation matrix. 

Figure \ref{fig:mvp:dendro} shows many interesting relationships among the HCCs.
For example, a clear group is formed by the HCCs for aspiration and specified
bacterial pneumonias, septicemia and sepsis, and cardio-respiratory failure.
Examining these further, we found that the prevalence of these conditions in the
overall population were approximately 5\%, 16\%, 22\% and respectively, but when
we subset to individuals who had aspiration and specified bacterial pneumonias,
the rates of sepsis and cardio-respiratory failure increased to 49\% and 59\%,
respectively. Alternatively, if we restrict our attention to individuals who had
cardio-respiratory failure, the rates for sepsis and aspiration and specified
bacterial pneumonias rise to 31\% and 13\%, respectively. 

Another potentially interesting cluster is the group containing cerebral
hemorrhage, major head injuries, coma, and seizure disorders. Upon
investigation, we found that having a condition that falls in the coma HCC has
an incidence of 1.5\% in the population overall, whereas the incidence rate
rises to 7.5\% for individuals with a diagnosis of seizure disorders or
convulsions. Additionally, the rates for cerebral hemorrhage were 2.3\% in the
general population and 8.5\% for those who had seizure disorders or convulsions,
while those for major head injury were 2.5\% and 7.7\%, respectively. 

While more investigation is needed regarding the causal relationships within the
two groups of HCCs above, we were able to apply the multivariate probit model to
a large data set to visualize the relationships between HCCs. This approach can
help practitioners identify groups for which intervention programs can be
created. For example, it would be worth investigating the potential of head
injury prevention programs for individuals with seizure disorders, or a program
focused on reducing the rate of sepsis in individuals with aspiration and
specified bacterial pneumonia. 

\subsection{Rural-Urban Divide in Specialty Care} \label{sec:mvp:rud}

In this section we try and understand how the utilization of specialists varies
across rural, urban, and suburban populations, while controlling for an
individual's chronic conditions. As in the previous section, we apply the
divide-and-conquer approach to a data set of Medicare Advantage enrollees. We
restrict our attention to continuously enrolled individuals from 2018-2019 to
get an accurate understanding of their chronic conditions and specialist
visitation patterns in a particular year. For predictors, we use an individual's
age, gender, HCCs in 2018, a flag for whether they were enrolled in a DSNP plan
at the start of 2019, and whether they lived in an urban, rural, or suburban
area at the start of 2019. We regress these variables on indicators for whether
or not an individual visited a particular specialty in 2019, and restrict the
dimension of the response variable to specialties that were visited by 0.5\% of
individuals. Our final dataset consists of $3.1$ million individuals, with $60$
specialties, and $84$ predictors. As with the previous analysis, we fit the
model using 40 factors.

\begin{figure}
	\captionsetup[subfigure]{justification=centering}
	\centering
	\begin{subfigure}{\textwidth}
		\includegraphics[scale=0.675]{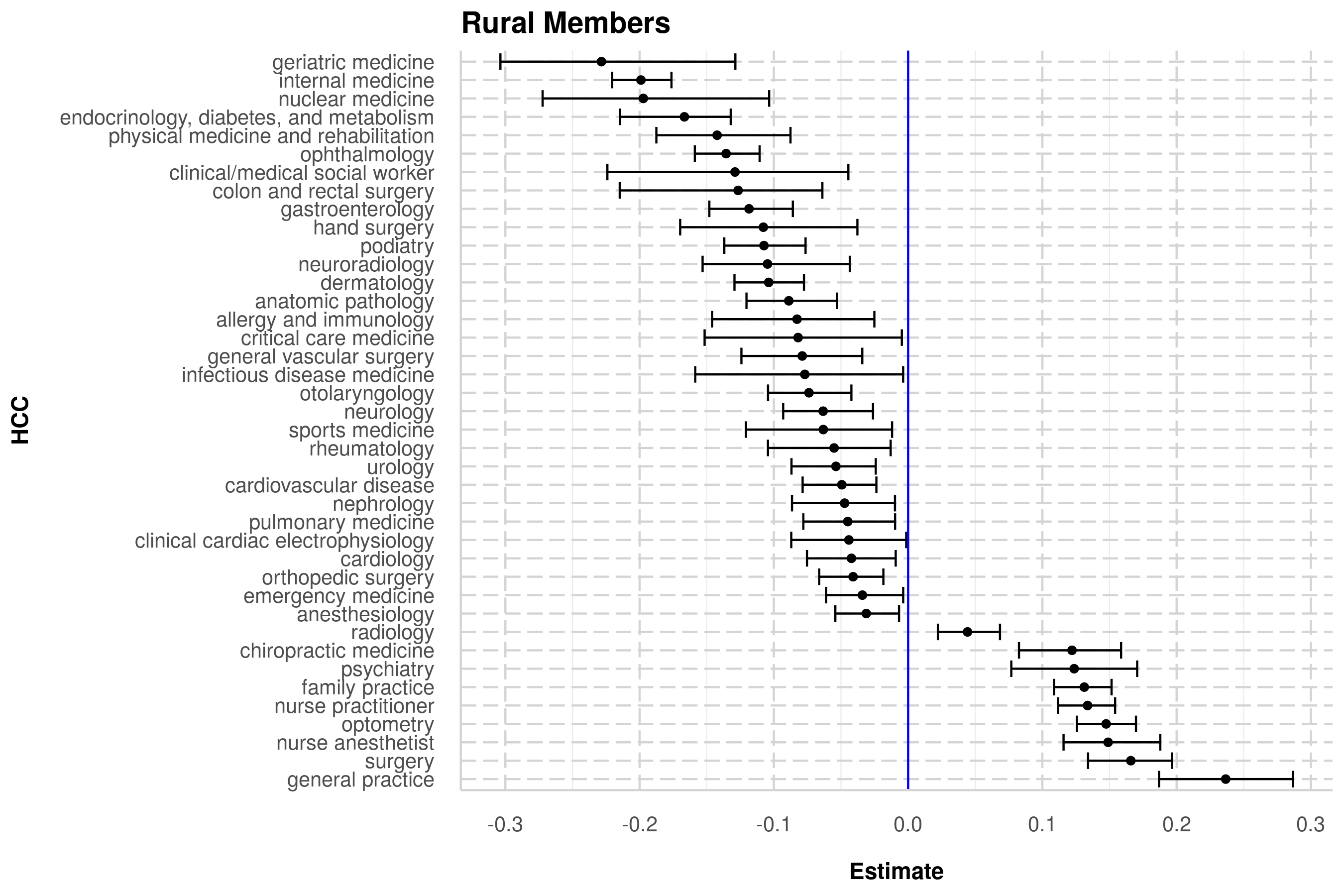}
	\end{subfigure}
	\begin{subfigure}{\textwidth}
		\includegraphics[scale=0.675]{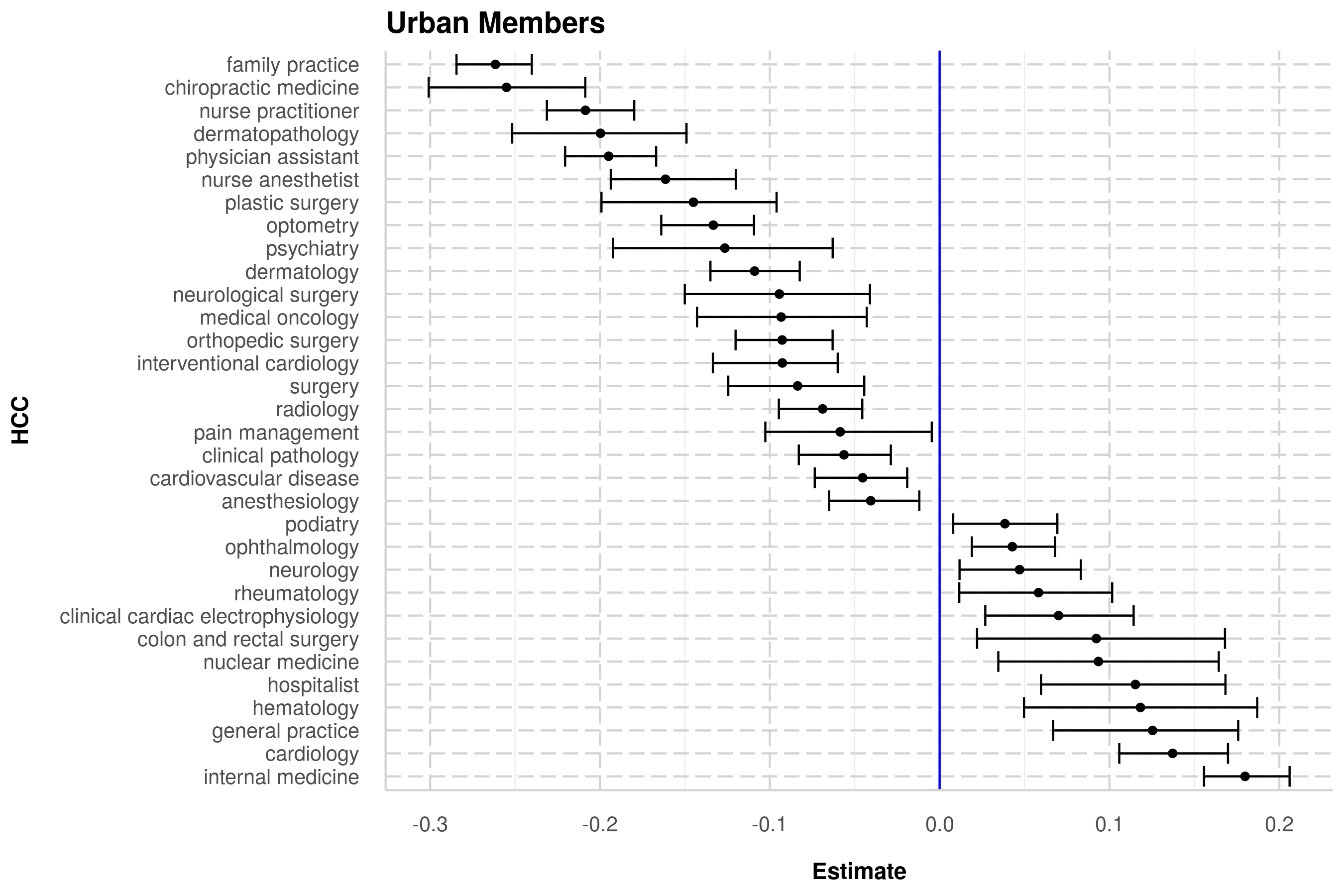}
	\end{subfigure}
  \caption{Specialties for which the 95\% credible interval for the urban or
  rural regression coefficient did not include zero.}	\label{fig:mvp:ur_signif}
\end{figure}

Figure \ref{fig:mvp:ur_signif} shows the specialties for which the 95\% credible
intervals for the urban and rural indicators did not contain zero, with suburban
residence used as the baseline. Estimated regression coefficients greater than
zero indicate that individuals in the group were more likely to visit a
particular specialty, relative to suburban individuals, after accounting for
their chronic conditions and demographics. It is interesting to note that many
meaningful specialties have negative coefficients for rural individuals. For
example, rural individuals were less likely to visit a doctor who specialized in
endocrinology, diabetes and metabolism despite its prevalence in the general
population. Additionally, rural individuals were less likely to visit
ophthalmologists but more likely to visit optometrists, which may point to a
mismatch of care since ophthalmologists can perform minor surgery for eye
conditions. On the other hand, rural individuals had a higher utilization of
nurse practitioners relative to suburban individuals; further investigation is
warranted into whether this difference exists due to gaps in specialist care,
and whether a larger than expected number of visits to nurse practitioners has a
detrimental impact on the quality of care. Finally, we also note that rural
individuals visited geriatric medicine specialists at a lower rate, which
highlights the lack of geriatric specialty availability in rural areas despite
the higher prevalence of older individuals \citep{hintenach2019training}.

While rural individuals visit nurse practitioners and family practice doctors at
higher rates than their suburban counterparts, urban individuals visit them at
relatively lower rates. Additionally, the relationship between ophthalmology and
optometry is flipped for urban individuals, and urban individuals visit internal
medicine physicians instead of those with family practice as their primary
specialty. Another interesting point of note is that, relative to their suburban
counterparts, urban individuals have a lower likelihood of visiting a physician
assistant or a nurse practitioner. This under-utilization may point to
opportunities for optimizing care, since nurse practitioners and physicians
assistants tend to be cheaper per visit than other specialties. Finally, it
should be noted that urban individuals visit endocrinologists at higher rates
than their suburban counterparts, while rural individuals visit them at lower
rates, which provides an actionable insight into a potential care gap between
rural and urban diabetic enrollees. 

\section{Conclusion} \label{sec:mvp:discuss}

In this paper, we show how a divide-and-conquer approach to MCMC can be utilized
to analyze large scale medical databases. We conducted a simulation study
comparing a stochastic variational algorithm with divide-and-conquer MCMC for
the multivariate probit model, and showed that the stochastic variational
approach underperforms in regards to estimation of the correlation matrix. We
also compared the performance of two different posterior combination algorithms
to understand their behavior as shard size remains fixed and the number of
samples increases. Our approach of combining divide-and-conquer MCMC with the
multivariate probit model drastically reduces the run time of the multivariate
probit Gibbs sampler when applied to a large dataset. 

In our applications, we explored two important questions for the health care
system. First, we found patterns of multimorbidity that can be used to create
health care programs tailored to specific disease profiles. Second, we
investigated how utilization of particular specialties differs between urban,
suburban, and rural individuals, and found that a potential gap in care exists
for diabetic individuals. A limitation of our analysis is that we fixed the
number of factors in our MCMC algorithms; a full Bayesian analysis would require
us to use model averaging to incorporate results from a differing number of
factors. However, it is unclear how to conduct model averaging in the context
of a divide-and-conquer approach, since it is an open question as to whether
model averaging be conducted at the shard level or at the combined posterior
level. We posit that the appropriate level to combine different models would be
the individual shard level, which would allow the optimal number of factors to
vary by subsample. A detailed study of model averaging in the context of
divide-and-conquer MCMC is beyond the scope of this work, and we leave the
question to be addressed by future research. 

While we focused our attention on analyzing multivariate binary datasets, we
hope that future research uses the latent factor model framework with
divide-and-conquer MCMC to extend our approach to multivariate count data and
multivariate mixed (binary, count, continuous) data. For example, it would be
interesting to analyze the number of visits to different specialties, or to
simultaneously predict the number of hospitalizations, the onset of new
diagnoses, and total cost. Finally, we also aim to incorporate a greater number
of predictors into our analyses (lab results, medications), and leverage GPU
clusters to conduct variable selection on large scale medical databases. 

%

\bibliography{references}
\bibliographystyle{plainnat}

\end{document}